# Interferon-induced transmembrane protein 3 (IFITM3) and its antiviral activity

Jimenez-Munguia I.,[1,†] Beaven A.H.,[2,3,†] Blank P.S.,[1] Sodt A.J.,[2*] Zimmerberg J.[1*]

[1] Section on Integrative Biophysics; Division of Basic and Translational Biophysics, *Eunice Kennedy Shriver* National Institute of Child Health and Human Development (NICHD), National Institutes of Health (NIH);

[2] Unit on Membrane Chemical Physics; Division of Basic and Translational Biophysics, *Eunice Kennedy Shriver* National Institute of Child Health and Human Development (NICHD), National Institutes of Health (NIH); and

[3] Postdoctoral Research Associate Program, National Institute of General Medical Sciences; National Institutes of Health (NIH), Bethesda, MD 20892, USA.

[†] Equal contribution.

[*]Corresponding Authors

## *Abstract*

Enveloped viral infections require fusion with cellular membranes for viral genome entry, occurring only following interaction of viral and cellular membranes allowing fusion pore formation, by which the virus accesses the cytoplasm. Here, we focus on interferon-induced transmembrane protein 3 (IFITM3) and its antiviral activity. IFITM3 is predicted to block or stall viral fusion at an intermediate state, causing viral propagation to fail. After introducing IFITM3, we describe the generalized lipid membrane fusion pathway and how it can be stalled, particularly with respect to IFITM3, and current questions regarding IFITM3's topology. Specific emphasis is placed on IFITM3's amphipathic α-helix (AAH) $_{59}$V-$_{68}$M, necessary for antiviral activity. Calculations are reported of hydrophobicity and hydrophobic moment of this peptide and active site peptides from other membrane-remodeling proteins. Finally, we discuss the effects of post-translational modifications and localization, how IFITM3's AAH may block viral fusion, and possible ramifications of membrane composition.

## *Manuscript*

Globalization favors the spread of novel viruses worldwide. Indeed, several viral outbreaks have occurred in the 21st century including influenza A (H1/N1)pdm09, Severe Acute Respiratory Syndrome (SARS), and most recently, coronavirus disease 2019 (COVID-19) [1]. Considering the rate of viral mutation, the acquired immune response is insufficient for preventing virus spread. Viral infections create a significant, negative impact on the healthcare system. Globally, four million deaths annually are associated with acute lower respiratory tract infections and in pandemic years the flu is a top human killer [2]. Influenza virus, even in non-pandemic years, is responsible for killing 250,000–500,000 people globally with an associated annual cost ranging from 71–167 billion USD [2]. The spectre of more frequent and prolonged viral pandemics, like COVID-19, highlight the urgency for advanced research and development of pan-viral therapeutics.

Human immune response against viral infections involves the activation of interferon (IFN) signaling [3]. Viral binding to pattern recognition receptors (PRRs) trigger the synthesis and secretion of IFNs, which induces IFN-stimulated genes (ISGs) [3] leading to synthesis of ~2000 proteins. Specifically, IFN type I ($\alpha$) and II ($\gamma$) stimulate the production of the interferon-induced transmembrane protein family (IFITMs) [3]. The human IFITM protein family consists of five members (IFITM1, 2, 3, 5, and 10) located on chromosome 11[4]. IFITM3 is potent against several enveloped viruses including Dengue, Influenza A (IAV) H1N1, Zika, coronavirus, hepatitis C, West Nile Virus, vesicular stomatitis virus (VSV) and human immunodeficiency virus (HIV), SARS-CoV2 [5,6].

Enveloped viral infection is mediated by fusion of the viral envelope to either the plasma membrane or endosomal membrane (after update by endocytosis). For viral-endosomal membrane fusion, it is triggered by some combination of low pH, receptor binding in acidic conditions, and the action of protease [7]. How IFITM stabilizes cellular membranes against viral envelope fusion is not well defined; even the stage in which IFITM interferes is contradictory [8,9]. Here, the recently proposed mechanisms for IFITM3's antiviral activity will be evaluated, specifically how IFITM3 interacts with membrane lipids to impair viral-host membrane fusion and thus viral cell entry [10,11]. After introducing IFITM3, as well as the generalized lipid membrane fusion pathway, this review will summarize early work on the topology and secondary structure of IFITM3, delineating attempts to isolate the membrane-interacting parts of IFITM3. This will culminate in a detailed discussion of IFITM3's amphipathic $\alpha$-helix region $_{59}$V-$_{68}$M; hereafter referred to as IFITM3's AAH. The next section reviews recent work linking IFITM3's AAH activity to the topology, localization, and aggregation of IFITM3. The impact of post-translational modifications on IFITM3's activity (specifically *S*-palmitoylation that induces membrane remodeling) follows. Finally, the role of membrane cholesterol will be discussed.

***The interferon-induced transmembrane protein 3***. IFITM3 has been implicated in a number of cellular functions including cell adhesion, stem cell migration, and leucocyte apoptosis [12]. Ubiquitously expressed in endothelial cells [13], its intracellular localization is mainly at endosomal and lysosomal membranes, although it is identified also in plasma and endoplasmic reticulum membranes in unstimulated cells with further distribution in membrane-derived vesicles after IFN stimulation [14,15]. IFITM3 is active against viruses that enter cells via pH-dependent or pH-independent fusion (e.g., influenza A or HIV, respectively) in the endosomal pathway [16]. That is, the antiviral activity of IFITM3 against enveloped viruses is exerted at the early stages of virus replication – after virus binds to the cell's surface but prior to viral mRNA production [14]. In the endosomal pathway, IFITM3 does not affect viral binding to surface receptors, virus endocytosis, or trafficking of viruses to late endosomes [14]. Instead, the loss of viral RNA in acidic inclusions of IFITM3-producing cells infected with H1N1 suggests virus entrapment in acidic compartments with further lysosomal degradation [14]. This implies that IFITM3 interferes with the virus-endosomal membrane fusion process irrespective of receptor binding and endosomal trafficking [9,17,18].

***The lipid membrane fusion pathway***. Many enveloped viruses fuse their membranes with endosomal membranes following endocytosis, releasing their genome into the cell for replication (Figure 1) [15,19–22]. The energy barrier for fusion of symmetric, model lipid membranes is estimated to be in the 10s to 100s of $k_BT$ [23,24], meaning that efficient biological fusion must be mediated by viral proteins that bring the viral and endosomal membranes near to contact to lower the energetic barrier [25,26]. Yet, the barrier varies with lipid composition. Cellular plasma membranes, from which endosomes and some viral envelopes emerge, are highly asymmetric in lipid saturation, head group chemistry, and charge; composition varies further between the membranes of organelles [27,28].

For viral fusion, the outer leaflet of the viral membrane and inner leaflet of the endosomal membranes merge first. Where first there were four distinct leaflets (i.e., viral inner, viral outer, endosomal inner, and endosomal outer), there are now *three* distinct leaflets: i) the viral inner leaflet (Figure 1, region 4(b)); ii) the mixed leaflet (Figure 1, regions 4(c, d)); iii) and the outer leaflet of the endosome (Figure 1, regions 4(a, e, f)). In two-dimensions, curvature is considered *negative* if a leaflet is bent toward its head groups. Conversely, curvature is *positive* if a leaflet is bent toward its tails. In three-dimensions, a particular region can be both negatively *and* positively curved, e.g., on a saddle (Gaussian curvature). For example, both the stalk and diaphragm have negative curvature in the bowl region near the merge point, region 4(e). Moving laterally on the surface from the merge point, curvature becomes positive in the viewing plane and remains negative out of the plane (see region 4(f)).

At hemifusion, there is no substantial cellular uptake of viral material. Only viral lipid content has mixed. If stalled at hemifusion, the endosome will traffick to the lysosome and degrade the virus – preventing replication. Thus, cellular antiviral proteins can stall the pathway at hemifusion by shifting leaflet curvature preference [29,30]. For example, lipids with positive monolayer intrinsic curvature (e.g., lyso-phosphatidylcholine) inhibit hemifusion when added to the proximal leaflets (e.g., virus outer and endosome inner) and facilitate pore formation when added to distal leaflets (e.g., virus inner and endosome outer). Conversely, lipids with negative monolayer intrinsic curvature (e.g., phosphatidylethanolamine) facilitate hemifusion when added to proximal leaflets and inhibit pore formation when added to distal leaflets.

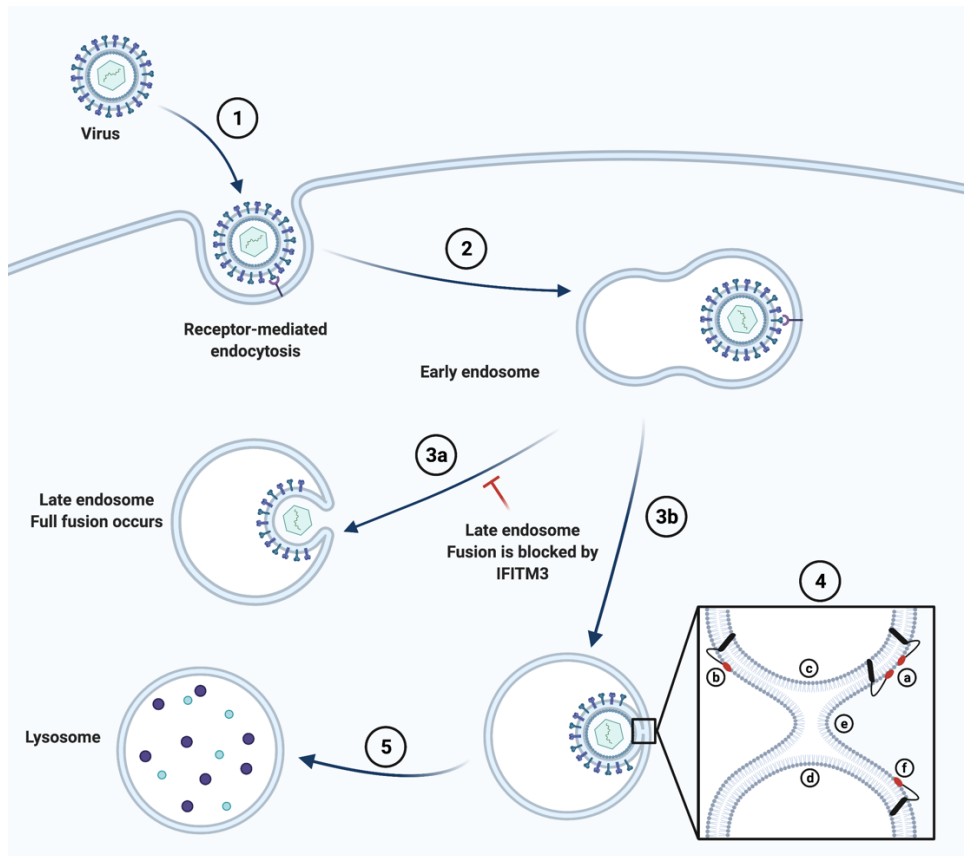

**Figure 1: Schematic representation of current ideas on enveloped viral entry and the blockage of viral entry by the IFITM3 amphipathic α-helix.** The viral entry route occurs as follow: 1) Virus is bound to the cell surface by a specific-receptor recognition. 2) Virus experiences receptor-mediated endocytosis. 3a) Virus is trafficked from the early endosomes to the late endosomes, during which time acidification of endosome triggers the activation of the virus fusion protein to induce fusion between the virus and endosome membranes. Fusion pore formation allows viral genome release into the cell's cytoplasm for further initiation of viral replication in the nuclei. Viral entry is blocked by IFITM3 by prevention of virus release from late endosomes (red T). 3b) Virus is trapped in late endosomes at the hemifusion stage (4) in which the fusion pore formation is prevented by the amphipathic helix (AH) of IFITM3 (black rectangle). The leaflet associated with markers (a), (e), and (f) is the endosome outer leaflet *only*. The leaflet associated with marker (b) is the viral envelope's inner leaflet *only*. The hemifused leaflet joining endosome inner leaflet with viral envelope outer leaflet is marked by (c). Also, note that (a) IFITM3 can aggregate or work alone to block the fusion pathway, and 4b) IFITM3 could work by its incorporation into the viral envelope.

*The IFITM3 dilemma in the fusion pathway*. Is it possible that IFITM3 acts like an aforementioned lipid that inhibits fusion? While IFITM3's membrane fusion disruption is implicated experimentally [10,31–37], but the (1) stage and (2) mechanism by which IFITM3 interferes in the fusion process is unclear.

(1) Whereas *cell-cell fusion* is inhibited before hemifusion [8] (viral fusion proteins expressed on the plasma membrane), viral fusion *with endosomes* is blocked after hemifusion [9]. This is a radical difference – the cell-cell fusion assay that evaluates plasma membrane-plasma membrane fusion shows IFITM-dependent blockage of fusion at hemifusion [8], but when fusion is evaluated in the context of viral infections (e.g. endosomes containing viral particles), hemifusion between the viral and endosomal membranes occurs [9] with subsequent blockage either of fusion pore formation or expansion.

(2) In both experiments, the authors invoke IFITM3-dependent alteration of membrane curvature to explain the stage at which fusion is blocked, but each group invokes the opposite cuvature [8–10]! For endosomal fusion, the antifusion activity of IFITM3 to its AAH [38]. Clearly, it is imperative that experiments and modeling be performed to determine if the IFITM3's AAH produces positive [8] or negative [11] leaflet intrinsic curvature.

*IFITM3 topology, secondary structure, and their putative roles in blocking fusion*. To evaluate IFITM3's mechanism on a specific leaflet, we must know how the protein and its membrane domains are oriented in the bilayer. However, IFITM3's 133 residues can be distributed across the membrane in a variety of topologies (Figure 2).

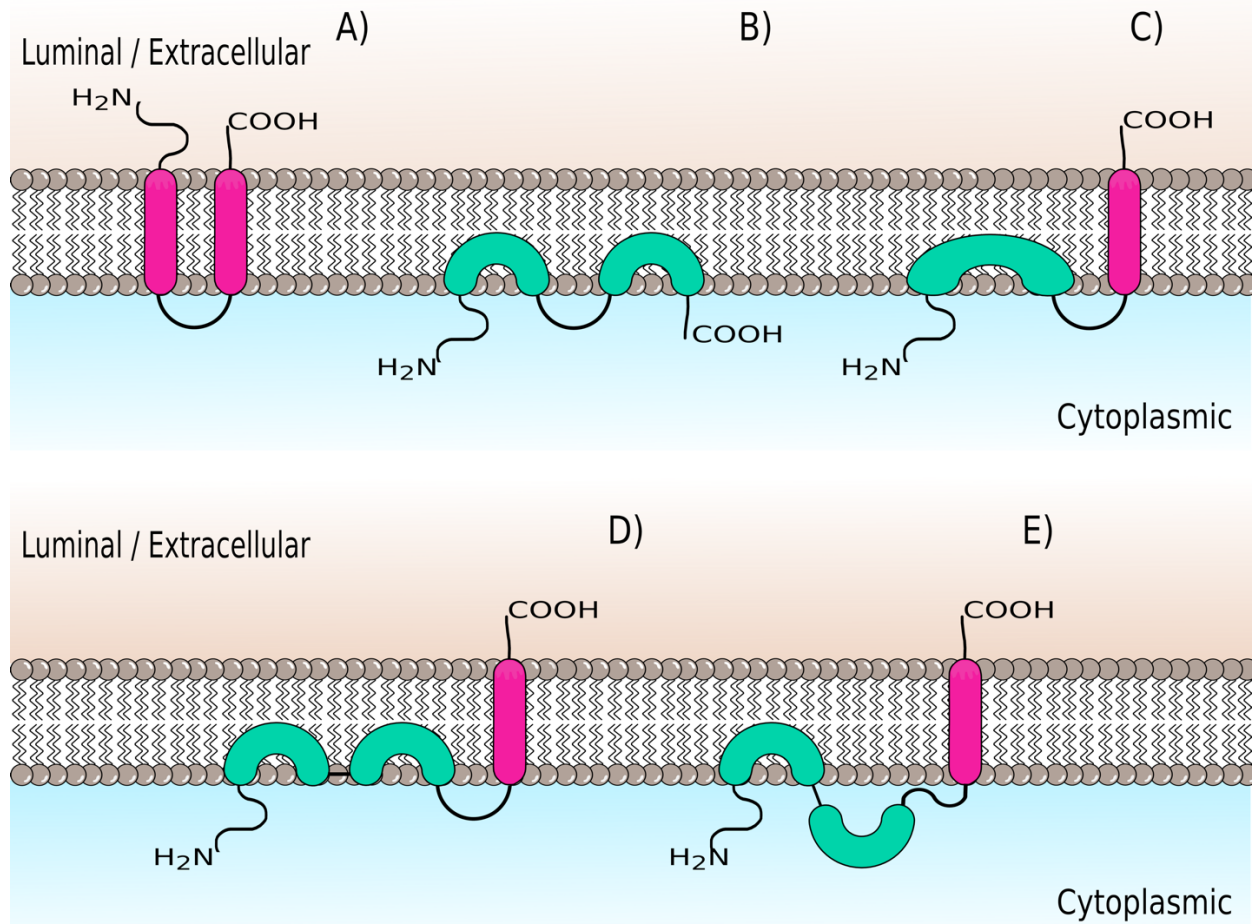

**Figure 2: Predicted topologies for IFITM3.** A) Two pass transmembrane protein with its *N*- and *C*-terminus exposed in the endosome lumen (extracellularly) [15]. B) An intramembrane protein with its *N*- and *C*-terminus exposed in the cytoplasm [16,32]. C) Type II transmembrane protein with its N-terminus exposed in the cytoplasm and its *C*-terminus exposed in the lumen [39]. D) A single long transmembrane helix with its *N*-terminus exposed in the cytoplasm, its *C*-terminus exposed in the lumen, and an intramembrane segment in the cytoplasmic leaflet (corresponding to two regions detected by NMR that interact with the membrane) [40]. E) As in (D) but with the middle α-helix in the cytoplasm. For linear sequence and secondary structure determinations, see Figure 1A of Chesarino et al. [38].

Indeed, IFITM3's topology in the membrane may be uncertain because sub-populations of IFITM3 could shift orientation upon specific stimuli: biochemical evidence appears to contradict a single topology [39,41,42]. Additionally, topologies C–E require long transmembrane domains.[10] For example, the structure predicted in Chesarino et al. suggests a transmembrane domain that is ~58.6 Å long (each residue contributes 1.5 Å rise); nearly twice the length of a prototypical membrane thickness of ~30 Å. Indeed, all-atom MD simulations of IFITM3 displayed severe TM tilt angles [43]. The unusual length of the domain may indicate that this feature is less topologically stable than common TM lengths. Nevertheless, it is likely that IFITM3 has a *C*-terminus facing the lumen, a transmembrane section, and an intramembrane portion at the *N*-terminus (this is true in the four most recently proposed topologies). Already these features constrain hypotheses of how IFITM3 prevents viral infections by blocking fusion. For example, topologies B–E have α-helices predicted to be embedded in the cytoplasmic leaflet. The leaflet in which the AAHs lies is an important characteristic for determining how IFITM3 blocks fusion pores.

**BOX:**

***The amphipathic region $_{59}$V-$_{68}$M of IFITM3***. The hydrophobic moment of IFITM3's ($_{59}$V-$_{68}$M; see topology E, Figure 2) appears to be correlated with the antiviral activity of IFITM3, since mutating or deleting this region decreases IFITM3's antiviral efficacy [10,11]. To compare mutated and wild-type AAH to other amphipathic helices, we replicated the hydrophobic moment calculation of region $_{59}$V-$_{68}$M using HELIQUEST [44] (Figure 3 and Table 1). We highlight the 65F insertion mutation that has nearly the same hydrophobicity as wild type, but a distinctly different hydrophobic moment. The hydrophobic moment likely determines how anchored the peptide remains at the leaflet-water interface [45]. Strong anchoring above or below the leaflet's neutral surface would generate positive curvature or negative curvature, respectively. Therefore, we anticipate the hydrophobic moment to correlate with curvature generated by the peptide [46].

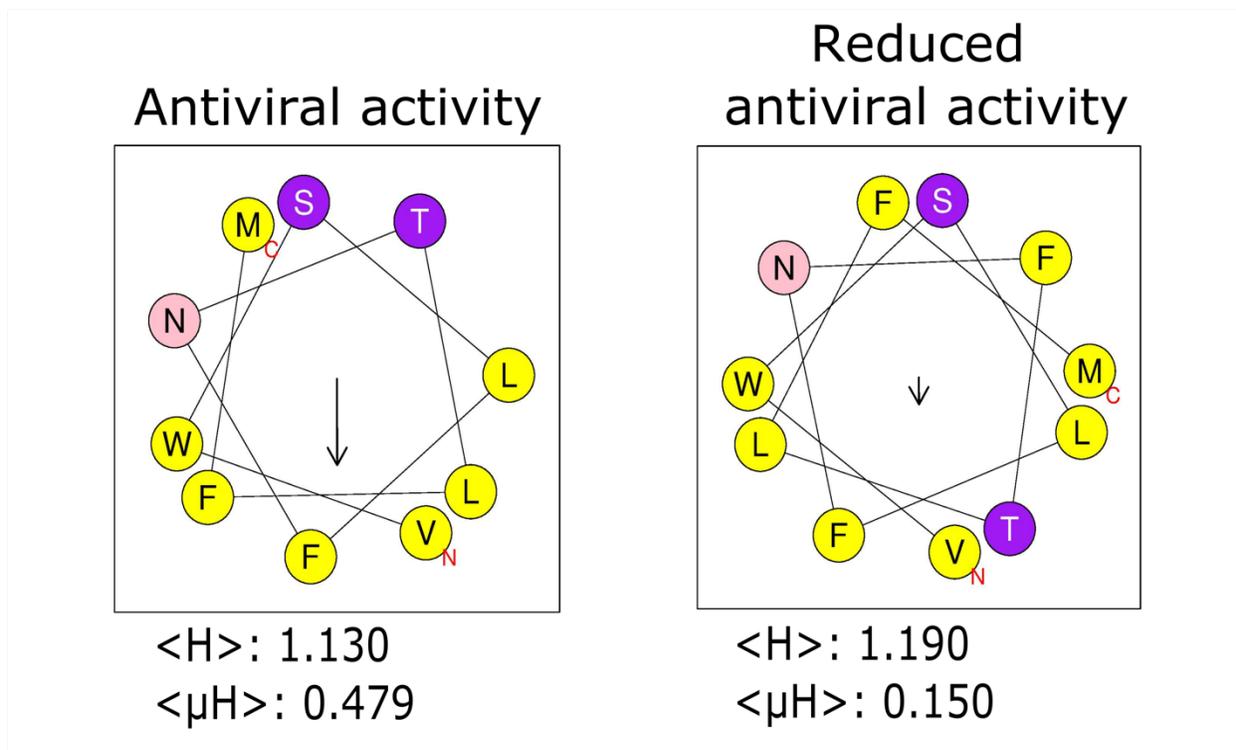

**Figure 3: Helical wheel representations of IFITM3's $_{59}$V–$_{68}$M region.** HELIQUEST [44] bioinformatic tool was used to plot the protein region $_{59}$V–$_{68}$M of IFITM3 with antiviral activity and the same protein region with a phenylalanine inserted in position 65 ($_{59}$V–$_{65}$F–$_{69}$M) with diminished antiviral activity. $\langle H \rangle$ and $\langle \mu H \rangle$ indicate hydrophobicity and hydrophobic moment values, respectively. The helical wheel for $_{59}$V–$_{68}$M with antiviral activity, and the inserted $_{65}$F version with diminished antiviral activity are shown. Hydrophilic and hydrophobic amino acids are colored in purple and yellow, respectively. Red letters indicate the *N*- and *C*- terminus. The arrow length and direction represents the hydrophobic moment. The theoretical range for hydrophobicity and hydrophobic moment are: $\langle H \rangle$ from 1.01 to 2.25 and $\langle \mu H \rangle$ from 0 to 3.26.

| Peptide | $\langle H \rangle$ | $\langle \mu H \rangle$ |
|---|---|---|
| **IFITM3 WT** (VWSLFNTLFM) | 1.130 | 0.479 |
| **IFITM3 mutant** (L62Q) | 0.938 | 0.519 |
| **IFITM3 mutant** (Insert 65F) | 1.190 | 0.150 |
| **IFITM3 mutant** (F63Q) | 0.929 | 0.282 |
| **A2-17 R10L/L11R** (LRKLRKRLLLRWKLRKR) | 0.143 | 0.329 |
| **A2-17 L14R/R15L** (LRKLRKRLLRLWKRLKR) | 0.143 | 0.788 |
| **ARFGAP1** (PPPQKKEDDFLNNAMSSLYSGWSSFTTGASRFAS) | 0.293 | 0.345 |
| **divIVA** (EVNEFLAQVRKDYEIVLR) | 0.300 | 0.480 |
| **Piscidin 1** (FFHHIFRGIVHVGKTIHRLVT) | 0.674 | 0.584 |
| **Piscidin 3** (FIHHIFRGIVHAGRSIGRFLT) | 0.638 | 0.580 |

**Table 1.** Hydrophobicity $\langle H \rangle$ and hydrophobic moment $\langle \mu H \rangle$ of peptides determined by HELIQUEST [44]. HELIQUEST was also used by Chesarino et al. (see Table EV1 in that work) [10]. The hydrophobic moments (theoretical range of 0 to 3.26) for IFITM3's $_{59}$V–$_{68}$M and mutants are compared to arginine-rich cell-penetrating peptides (A2-17) [46], the ALPS1 domain of human ARFGAP1 [47–49], the predicted AAH region of the bacterial division factor of *Bacillus subtilis*' divIVA [50,51], and piscidins 1 and 3 [52]. The IFITM3 data matches with that published in Chesarino et al. [38] and the trends for A2-17 match Takechi-Haraya et al. [46] (note that A2-17 L14R/R15L increases the duration and charge flux from stable membrane pores compared to A2-17 R10L/L11R). The remaining data are new determinations. It is difficult to make comparisons across studies given methodology differences, but within studies, we can conclude that: i) A2-17 L14R/R15L has a larger hydrophobic moment and induces stronger *positive* curvature than A2-17 R10L/L11R [46]; ii) divIVA has a larger hydrophobic moment but induces *similar positive* curvature to AFRGAP1 [53]; and iii) piscidin 1 and 3 induce *similar positive* curvature (in POPC:cholesterol, 4:1 mol%) and have very similar hydrophobic moments [52]. As implied by Chesarino et al. [38], it appears that hydrophobic moment is a reasonable, but imperfect, indicator of curvature induction.

***The impact of post-translational modifications on IFITM3 activity***. IFITM3's antiviral activity is reduced by some PTMs, including lysine methylation ($_{88}$K), tyrosine phosphorylation ($_{20}$Y) and ubiquitination (Figure 4) [32,34,35]. Conversely, mono-methylation on $_{88}$K is induced during infections caused by Vesicular stomatitis virus (VSV) and IAV [54] but demethylation of $_{88}$K by histone demethylase LSD1 prevents RNA virus replication, and impairs viral infection [55]. Furthermore, phosphorylating $_{20}$Y blocks the endocytosis signal motif YXXΦ (of which $_{20}$Y is part) leading to a redistribution and accumulation of IFITM3 in the plasma membrane [34,41].

Contrapositively, *S*-palmitoylation helps to anchor IFITM3 to the membrane and promotes its antiviral activity [31]. *S*-palmitoylation frequently occurs on cysteines adjacent to transmembrane domains, inducing transmembrane domain tilting, and producing weak negative curvature in the embedded leaflet [56,57]. Indeed, the first all-atom MD simulations of IFITM3 displayed severe TM tilt angles with AAHs bound more strongly to the membrane when *S*-palmitoylation is present [43]. Defects in IFITM3's *S*-palmitoylation lead to a different phenotype with less affinity for the membrane and inefficient antiviral activity [32]. In IFITM3, $_{71-72}$C and $_{105}$C are susceptible to palmitoylation. Considering the close proximity of $_{71-72}$C to AAH $_{59}$V-$_{68}$M, palmitoylation on $_{71-72}$C could facilitate $_{59}$V-$_{68}$M insertion into the cytoplasmic leaflet of endosomal membranes by increasing the hydrophobicity in the proximal membrane region, stabilizing IFITM3 insertion into endosomal membranes; and consequently, cause inhibition of fusion pore formation.

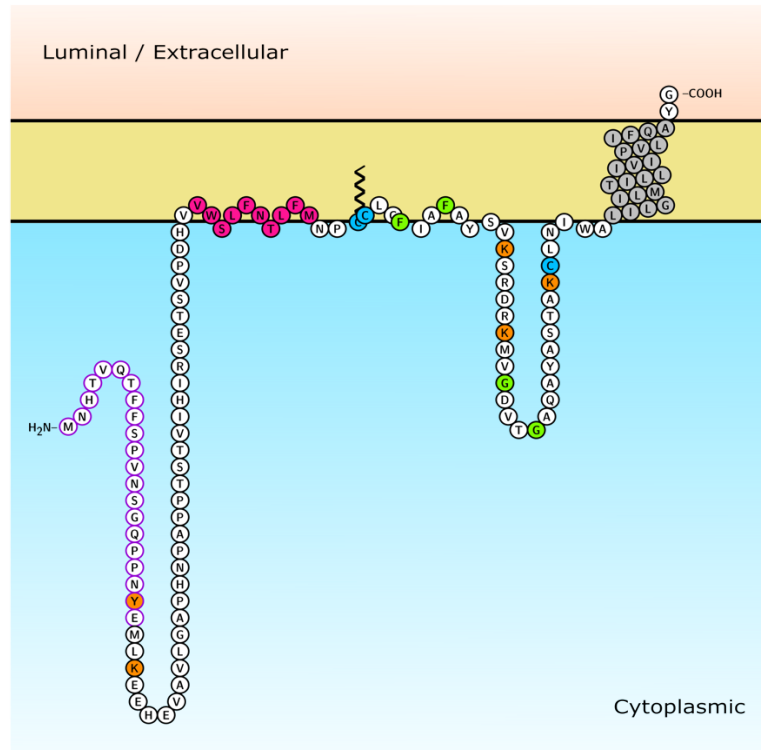

**Figure 4: Regions of the IFITM3 protein sequence involved in the performance of IFITM3.**
A) Amino acids described in the literature that are important for the activity of IFITM3. The *N*-terminus (residues 1-21), which is critical for IFITM3's endosomal localization, is highlighted in purple [16]. In green are two phenylalanine residues ($_{75}$F and $_{78}$F) necessary for the antiviral activity of IFITM3 [58] and two glycine residues ($_{91}$G and $_{95}$G) that represent a GxxxG motif necessary for the IFITM3 dimerization and pore formation impediment [59]. Residues highlighted in orange are post-translational modified amino acids that negatively impact IFITM3 performance including phosphorylated $_{20}$Y [16], ubiquitinated $_{24}$K, $_{83}$K, and $_{104}$K [32], and the ubiquitinated and methylated $_{88}$K [32,54]. In blue are *S*-palmitoylated cysteines ($_{71-72}$C and $_{105}$C) that positively impact IFITM3 anti-viral performance [60]. The transmembrane domain is colored gray and the AAH ($_{59}$V-$_{68}$M) is colored red.

***The role of membrane composition.*** Given that IFITM3 production occurs concurrently with aberrant late endosomal cholesterol accumulation, sterol concentration itself may play a role in the early antiviral defense [61,62]. For example, the interaction between IFITM3's AAH and the endosomal membrane might be influenced by an increase in the lipid order caused by the accumulation of cholesterol in late endosomes in IFITM3-expresing cells [61,62]. Additionally, negative Gaussian curvature, which can occur at the necks of fusion pores, has negative mean curvature in both leaflets, leading to a net thinned membrane interior. This thinning might influence sterol distributions around the pore – affecting pore formation energetics.

Supporting this hypothesis, recent work has shown direct evidence that sterols do affect energetic barriers associated with the pore narrowing/widening step in endo-/exocytosis [63,64], and a substantial body of work has focused on altering the host's cholesterol metabolism to inhibit enveloped viral replication [65–70]. Consequently, a hypothesis suggests that each virus has a preferred plasma membrane or endosomal cholesterol content that can affect when and where it targets optimal replication. Recent experiments and simulations have demonstrated that IFITM3's *S*-palmitoylation interacts directly with endosomal cholesterol, and the quantity of these interactions (i.e., more/less cholesterol or more/less palmitoylation) modulates antiviral activity against IAV, SARS-CoV-2 and EBOV [42]. That said, cholesterol's importance relative to IFITM3 blockage of the hemifusion step is disputed, with experiments suggesting that only the IFITM3 protein is responsible for full fusion blockage [8,9,61,71]. Determining how peptides [72], cholesterol, and other lipids partition into a pore's saddle curvature, thinned/thickened leaflets, and net thinned membranes is of extreme interest for future experiments, simulations, and theory.

**Conclusion.** IFITM3 impairs viral infections caused by several enveloped viruses, for which membrane fusion is required for cell entry. IFITM3 significantly affects the progression of viral infections by preventing fusion and blocking viral entry into the cytoplasm [9,14]. Fusion is impaired by IFITM3's AAH, which allows viral-endosome hemifusion but not full pore formation. The sign of curvature induction by the AAH is a key parameter determining the stability of fusion and hemifusion structures. Additionally, post-translational modifications affect the activity of IFITM3. *S*-palmitoylation favors IFITM3 antiviral activity by facilitating interactions with and/or enhancing AAH insertion into the endosomal membrane or viral envelope. Modification of the AAH's hydrophobicity, by mutation or palmitoylation, leads to changes in IFITM3's antiviral activity [31,32,38]. Finally, we posit that endosome membrane composition is important for antiviral function, and that the endosomal membrane environment works synergistically with IFITM3. Research focused on the local action of IFITM3's AAH and its interactions with the membrane will contribute to a better understanding of the mechanisms involved in the impairment of viral infection, specifically in the putative prevention of the transition from hemifusion to pore formation.

*Acknowledgements.* This work was supported by the intramural program of the NICHD. AHB was supported by a Postdoctoral Research Associate (PRAT) fellowship from the National Institute of General Medical Sciences (NIGMS), award number 1Fi2GM137844-01. Figure 1 was drawn on biorender.com, and Figure 4 was created with the Protter webtool and Inkscape.